\documentclass[prl,twocolumn,showpacs,preprintnumbers,amsmath,amssymb]{revtex4}
\usepackage{amsfonts}
\usepackage{graphicx}
\usepackage{dcolumn}
\usepackage{float}
\usepackage{bm}

\newcommand{\com}[1]{}

\begin{document}

\title{Metastable states and macroscopic quantum tunneling in a cold atom Josephson ring}

\author{Dmitry Solenov\footnote{E-mail: solenov@lanl.gov} and Dmitry Mozyrsky\footnote{E-mail: mozyrsky@lanl.gov}}

\affiliation{Theoretical Division (T-4), Los Alamos National
Laboratory, Los Alamos, NM 87545, USA}

\date{\today}

\begin{abstract}
We study macroscopic properties of a system of weakly interacting neutral bosons
confined in a ring-shaped potential with a Josephson junction. We derive an effective
low energy action for this system and evaluate its properties. In particular we find that
the system possesses a set of metastable current-carrying states and evaluate the rates
of transitions between these states due to macroscopic quantum tunneling. Finally we
discuss signatures of different metastable states in the time-of-flight images and argue
that the effect is observable within currently available experimental technique.
\end{abstract}









\maketitle
Macroscopic Quantum Tunneling (MQT) is undoubtedly one of the most
remarkable manifestations of macroscopic quantum coherence, a
property intrinsic to superconductors and Bose-Einstein Condensate
(BEC) systems. The most well studied system exhibiting MQT is the so-called
phase qubit, a superconducting loop containing a Josephson junction, where MQT
occurs between metastable current-carrying states of the Cooper
pair condensate in the loop \cite{squid, weiss}. The main mechanisms controlling these
metastable states and the MQT arise from an interplay between
several factors, such as Josephson, inductive, and charging
energies, as well as dissipation due to shunting resistances
and/or quasiparticle excitations. The Josephson and the inductive
energies form an effective potential for the flux-phase variable
(i.e., the phase difference across the Josephson junction), while
the charging energy (resulting from plasma oscillations due to the
Coulomb interactions between electrons) gives rise to the
effective kinetic energy for the phase \cite{weiss, amb}. Dissipation is
also known to play a critical role in MQT dynamics, leading to a
substantial slow-down in tunneling rates \cite{CaldeiraLeggett,
DSDM} as well as to the localization of the magnetic flux in the
case of symmetric double well effective potential (realizable by
applying a $1/2$ magnetic flux quantum through the loop)
\cite{Leggrmp, shnir}.

In this Letter we consider dynamics and properties of a phase
qubit-like (Josephson ring) device based on a BEC of neutral
particles. Contrary to the conventional superconductor-based
devices, these systems did not receive much attention in
literature, - presumably due to lack of their experimental
realizations in the past. Our interest in such systems, however,
is {\it not} purely academic: advances in cold atom trapping, such
as the recently demonstrated ``painted potential" technique
\cite{Boshier}, open a clear path for the realization and
development of such phase qubits. As we will show below, these
cold atom persistent current devices posses a number of properties
similar to their superconducting counterparts. These properties,
however, are controlled by rather different physical mechanisms,
which rely not on particles' charge (like inductive and charging
energies of the in superconducting devices), but on the effects of
quantization due to the finite size of the system. Indeed, while
for superconductors the phase (Goldstone) mode is gapped due to
plasmon oscillations, it is gapless for the neutral particles
interacting through a short range potential. As a result, the
phase dynamics for the latter systems is susceptible to the
long-range fluctuations and is strongly affected by the system's
dimensionality, geometry of the traps, etc.
\begin{figure}
\includegraphics[width=6.5cm]{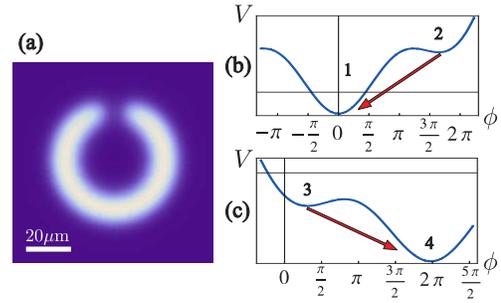}
\caption{(Color online) (a) The system of interest: Density profile of
a BEC in a "painted potential", i.e., a ring
with a tunnel junction (see Ref.~\cite{Boshier} for comparison
with actual experimental photographs). (b) and (c) Effective
potentials as a functions of phases across the Josephson junctions
for two cases: (b) $\Omega=0$, $\alpha=5.8$ and (c) $\Omega\neq
0$, $\alpha=5.8$ and $\phi_0=5.8$.}\label{BEC.eps}
\end{figure}

We consider a BEC confined in a ring-shaped potential containing a single
cut, i.e., a tunnel (Josephson) junction, see Fig. 1(a). We derive an effective action
for the phase difference across the Josephson junction.
Dynamics described by this action strongly depends on the system's size (i.e.,
the ring's circumference),
and is dramatically different in the two limiting cases, i.e., for
small and large BEC rings. In the latter case dynamics of the
phase $\phi$ across the junction corresponds to that of an
overdamped particle, diffusing in a periodic potential (due to the
Josephson energy at the junction, see below). In the former case the
"phase-particle" is underdamped and moves in the potential, which
is a sum of Josephson energy ($\sim\cos{\phi}$) and effective
inductive energy ($\sim\phi^2$). The inductive energy term is of
quantum origin: It arises due the kinetic energy of the BEC
particles confined in the ring.
The competition between Josephson and the effective inductive energies gives rise
to the formation of metastable current-carrying states, i.e., with
$\phi\neq 0$. We obtain criteria for the formation of these states and
discuss means of their control and possibility of their realization in
cold atom systems. We analyze MQT rates from these states as a
function of system's parameters for stationary and for rotating potentials.
Finally we outline signatures of these current-carrying states in
time-of-flight experiments.

The Hamiltonian for the system of interest is given by
\begin{equation}\label{eq:hamiltonian}
H=\!\!\int_0^L \!\!\!dx\,
\hat\psi^{\dag}(x)\!\left[-\frac{\nabla^2\!\!\!}{2m}
+\frac{\lambda}{2}\hat\psi^\dag(x)\hat\psi(x)+V(x)\right]\!\hat\psi(x),
\end{equation}
where potential $V(x)$ describes the tunnel barrier and the system
is assumed to be one dimensional~\cite{comm3} with periodic
boundary conditions. Here and in the following we use units with
$\hbar=1$ and $k_B=1$, unless stated otherwise. We would like to derive an
effective low energy action for the system described by the
Hamiltonian in Eq.~(\ref{eq:hamiltonian}). This can be carried out
as follows. In the
absence of the potential barrier $V(x)$ this action is readily
obtained by representing Bose field operators in terms of
hydrodynamic variables as
$\psi=\sqrt{\rho+\delta\rho}\,e^{i\varphi}$, where $\rho=N/L$ and
$N$ is total number of atoms in the ring. In that case the
linearization with respect to the small density variations
$\delta\rho$ yields an effective phonon Lagrangian in the long
wavelength limit~\cite{landau}. In the phase representation it
reads
\begin{equation}\label{eq:phonon-lag}
{\cal L}_0 = \int_0^L dx {1\over
2\lambda}\left[(\partial_\tau\varphi)^2 +c^2
(\partial_x\varphi)^2\right],
\end{equation}
where $c$ is sound velocity for an equivalent infinite
system ($c  = (\rho\lambda/m)^{1/2}$ in the weak coupling limit).
Here and in the following we use imaginary (Matsubara) time representation.
The spectrum of the phonons is  discrete due to the finite size of the system, with energy
spacing between the adjacent levels $\Delta E=2\pi c/L$.
If the barrier is sufficiently high, the inclusion of the tunnel barrier potential $V(x)$
changes the functional dependence of the Lagrangian density on
$\varphi$ only locally. That is, the BEC Lagrangian density in the region
with $V(x)=0$ is still given by Eq.~(\ref{eq:phonon-lag}),
supplemented by the appropriate boundary conditions at the
barrier (see below). For the high potential barrier
transitions across the barrier can be described by the usual hopping
term $\sim[\psi^\dag(0)\psi(L)+\psi^\dag(L)\psi(0)]$ and therefore
the effective action can be cast in the form
\begin{equation}\label{eq:phonon-action}
S = \int d\tau\,\left[ {\cal L}_0-\rho t\cos{\phi(\tau)}\right],
\end{equation}
\noindent
where $\phi=\varphi(0)-\varphi(L)$ is the phase difference across the junction and
$t$ is the transmission coefficient. Obviously the value of $t$ depends
on the shape of the potential barrier as well as it accounts for the depletion
of the density near the barrier. Moreover, it may depend on the interactions between
particles and their density~\cite{Leggett}. However, if the height of the barrier $V_0$ is large
compared to the interaction energy $\lambda\rho$, this dependence can be neglected,
and, as a rough estimate, one can use the standard semiclassical expression for the
rectangular barrier, $t\sim (1/md)\exp{[-d(2mV_0)^{1/2}]}$, where $d$ is the width
of the barrier. Furthermore, Eqs.~(\ref{eq:phonon-lag}, \ref{eq:phonon-action}) are applicable
only when Josephson energy per particle $\rho t/N$ is small compared
to the level spacing $\Delta E$, i.e., for $c\gg t$; otherwise the presence of the barrier
significantly affects the quasiparticle (i.e. phonon) states inside the ring and the weak link
picture is no longer valid. In this weak link limit the
supercurrent at $x=0$ and $x=L$ is vanishingly small, and therefore Eq.~(\ref{eq:phonon-action}) must
be supplemented by the boundary conditions $\partial_x\varphi(0,\tau) =\partial_x\varphi(L,\tau)=0$.

The action given by Eqs.~(\ref{eq:phonon-lag},
\ref{eq:phonon-action}) can be reduced to a local action by integrating out
the $\varphi(x,\tau)$ field everywhere except $x=0$ and $x=L$.
This is accomplished by introducing the functional
$\delta$-function according to
\begin{equation}\label{eq:partition-Z}
Z = \prod\limits_\omega\int{\mathcal{D}\phi \mathcal{D}\varphi}
e^{ - S} \delta[\phi(\omega)-\varphi(0,\omega)+\varphi(L,\omega)],
\end{equation}
\noindent and using identity $\delta(a)=(2\pi)^{-1}\int dz
\exp{(iza)}$. The fields $\varphi$ and $z$ can then be
integrated out (note that in doing so one should expand field
$\varphi(x, \tau)$ in $\cos{(\pi n x/L)}$, thus satisfying the boundary conditions above).
After some calculation we obtain that the partition function takes the
form $Z=\int \mathcal{D}\phi \exp{(-S_{\rm eff})}$, where
\begin{equation}\label{eq:general-action}
S_{\rm eff} = \int\frac{d\omega}{2\pi} |\phi(\omega)|^2
\frac{c\,\omega}{4\lambda\tanh{\left(\frac{\omega L}{2c}\right)}}
-\rho t \!\!\int d\tau\cos\phi(\tau).
\end{equation}

Effective action in Eq.~(\ref{eq:general-action}) is similar to that for an
impurity in the Luttinger liquid ~\cite{Kane}. The first term of
the right-hand side is different due to the finite length of
the ring. Indeed, for sufficiently large ring the
first term is proportional to $|\omega|$ and the action is that of
a dissipative particle in a periodic potential. In the opposite
small ring limit, the first term in Eq.~(\ref{eq:general-action})
yields
\begin{equation}\label{eq:action-slow}
\int d\tau \left[(L/24\lambda)\dot\phi^2(\tau) +
(\rho/2mL)\phi^2(\tau)\right],
\end{equation}
\noindent where the effective kinetic energy term describes the
lowest BEC mode, that is, motion of the BEC as a whole, and the
last ``inductive'' term arises due to the kinetic energy of the BEC particles in
the ring. Together with the cosine term in
Eq.~(\ref{eq:general-action}) this $\phi^2$ term provides
effective potential energy for the ``phase particle''. This
potential may have one (at $\phi=0$) or more (at $\phi\neq 0$)
local minima. The $\phi\neq 0$ minima correspond to the metastable states carrying non-zero current.
The shape of such effective potential can be conveniently characterized in terms of
the dimensionless parameter $\alpha = mtL$. For $\alpha =\alpha_1\approx 4.60$
the first two metastable minima appear at $\phi\approx \pm 1.43\pi$. Two more metastable states
appear when $\alpha>\alpha_2\approx 10.95$, and so on. Furthermore, the effective potential can be
modified by rotating the "painted potential". As a result the potential energy of the superfluid
in the rotating reference frame changes by
$-M\Omega$, where $\Omega$ is angular frequency of rotation and $M$ is angular momentum of the
superfluid (in the stationary frame) \cite{landau}, and therefore the effective potential $V(\phi)$ can be cast in the form
\begin{equation}\label{eq:potential}
V^{\rm eff}(\phi) = \rho t\bigg[{(\phi-\phi_0)^2\over 2\alpha} - \cos{\phi}\bigg],
\end{equation}
where $\phi_0 = 2\pi I\Omega$ (here $I$ is the moment of inertia
of the superfluid per atom, $I\simeq mL^2$). Such $V^{\rm
eff}(\phi)$ is equivalent to that of a Josephson flux qubit, i.e.,
a superconducting loop with a Josephson junction in a magnetic
field \cite{shnir}. For $\phi_0=\pi$ effective potential $V^{\rm
eff}(\phi)$ is a symmetric double well, while for $\phi_0>\pi$ the
minimum close to $0$ becomes metastable and then, upon further
increase of $\Omega$, disappears completely for $\phi_0 >
\phi_0^{\rm
cr}=[\sqrt{\alpha^2\!\!-\!\!1}\!+\!|\!\cos^{\!-\!1}\!(-\!1\!/\alpha)|]/\alpha$.

Note that in Eq.~(\ref{eq:potential}) the particle density
controls only the overall strength of the effective potential, but
not its relative shape. The latter depends only on the parameter
$\alpha$, which is determined by the properties of the ``painted
potential'', but not by the number of particles in the system.
This fact is rather advantageous from the experimental point of
view: While the number of particles in the system typically obeys
Poissonian statistics and therefore it can significantly fluctuate
in different experimental runs, the form of the painted potential
(i.e., the value of parameter $\alpha$ determining the onset of
the metastability) is easy to control with sufficiently high
precision.

Transitions between different minima of the effective potential in Eq.~(\ref{eq:potential}) are driven by thermal and
quantum fluctuations. The former mechanism dominates at sufficiently high temperatures, exceeding frequency of small Josephson
oscillations in a given well. Note that in the metastable region, i.e. for $mtL\geq 1$, this frequency is of the order of
the spacing between quasiparticle energy levels, $\Delta E$. For the system of $^{87}$Rb atoms $L\simeq 0.2\, {\rm mm}$ and
$\rho^{3D}\sim 10^{12}\,{\rm cm}^{-3}$, e.g. Ref.~\cite{Boshier}, $T^\ast\sim\Delta E\sim 0.2\,{\rm nK}$. Thus, above $T^\ast$ the rate of transitions
is determined by the thermal activation rate $\sim\exp{(-V_0^{\rm eff}/T)}$, where $V_0^{\rm eff}$ is the height of the effective
potential barrier in Eq.~(\ref{eq:potential}). In the following we evaluate transition rates for two cases: (I) For a system in
the first metastable state, i.e., for $\alpha_1<\alpha<\alpha_2$ and $\Omega=0$, e.g. Fig. 1(b); (II) For the rotating system
($\Omega\neq 0$), when the initially global minimum at $\phi=0$ becomes metastable, i.e., for $\pi\le\phi_0 \leq \phi_0^{\rm cr}$,
e.g., Fig. 1(c). In the former case the rate of thermally activated transition $\Gamma_T$ is given by
\begin{equation}\label{eq:rate-thermal-1}
\ln{{\Gamma_0\over\Gamma_T}}={0.86\rho\over
mLT}(\alpha-\alpha_1)^{3/2},
\end{equation}
\noindent
while for the latter case
\begin{equation}\label{eq:rate-thermal-2}
\ln{{\Gamma_0^\prime\over\Gamma_T^\prime}}={1.89\rho\over
mLT}(\alpha^2-1)^{-1/4}(\phi_0^{\rm cr}-\phi_0)^{3/2}.
\end{equation}
\noindent For the order of magnitude estimate both pre-exponential
factors $\Gamma_0$ and $\Gamma_0^\prime$ can be taken as frequency
of small Josephson oscillations $\sim\Delta E$~\cite{Gor}, while
the height of the effective potential barrier $V_0^{\rm eff}\sim
\rho/(mL)$ in Eqs.~(\ref{eq:rate-thermal-1},
\ref{eq:rate-thermal-2}) is of the order of $10$ nK (here and in
the following we use the same values for the system's parameters
as above; see also Ref.~\cite{Boshier}). Thus we expect that
thermally activated transition rates of the order of $1-10$ Hz are
readily observable at temperatures below $10$ nK.

For temperatures below $T^\ast\sim 0.2$ nK thermal activation
mechanism becomes ineffective and transitions are driven by
quantum fluctuations, i.e., by the MQT. The rate of the MQT can be
computed by evaluating Euclidean action over the classical
instanton trajectory $\Gamma = \Gamma_0
\exp[-S(\phi_{cl})]$~\cite{Coleman}. Measurable MQT rates
can be found near the spinodal instabilities of the metastable states,
i.e., when the effective potential barriers are relatively small. In this
regime one can expand the effective potentials around the local minima (state $2$ in
Fig. 1(b), e.g. case (I), and state $3$ in Fig. 1(c), case (II))
up to the cubic terms in ($\phi-\phi_{{\rm min}}$). The kinetic
energy, i.e., the first term in Eq.~(\ref{eq:general-action})
minus its value at $\omega =0$ (the second term in
Eq.~(\ref{eq:action-slow})), is non-local in Matsubara time $\tau$ and
therefore is difficult to be dealt with exactly. For,
$\alpha_1-\alpha\ll 1$, however, the instanton trajectory
$\phi_{cl}(\tau)$ contains only small frequencies. In this limit
the kinetic energy is local, e.g., the first term in Eq.
~(\ref{eq:action-slow}). In the non-local $\alpha_1-\alpha\geq 1$
regime we use variational approach by choosing
$\phi_{cl}(\tau)=\phi_{cl}^0\exp{(-|\tau|/\Delta\tau)}$, where
$\phi_{cl}^0$ and $\Delta\tau$ are variational parameters. The
results for $\Omega = 0$, i.e., case (I), are
\begin{equation}\label{eq:relax-rate}
\ln{{\Gamma_0
\over\Gamma_Q}}\!=\eta\,\left\{\!\!{\begin{array}{*{20}c}
   {2.07(\alpha-\alpha_1)^{5/4}}&{\alpha-\alpha_1\ll 1}\\
   {\!\!\!\!\!\!1.63(\alpha-\alpha_1)}&{\alpha-\alpha_1\geq 1}\\
\end{array}} \right. \!\!\!,
\end{equation}
\noindent where $\eta = \sqrt{\rho/(m\lambda)}$ and the
pre-exponential factor $\Gamma_0$ is again of the order of $\Delta
E$~\cite{comm5}. Evaluation of the MQT for the rotating system,
i.e., case (II), gives
\begin{equation}\label{eq:relax-rate2}
\ln\!\frac{\Gamma'_0}{\Gamma'_Q}\! =\eta\,\left\{\!\!
{\begin{array}{*{20}c}
   {\!1.65(\alpha^2\!\!-\!\!1\!)^{-3\!/8}\!(\phi_0^{\rm cr}-\phi_0)^{5\!/4}}&{\phi_0^{\rm cr}-\phi_0\ll 1}\\
   {\!\!\!\!\!\!\!\!1.88(\alpha^2\!\!-\!\!1\!)^{-9\!/8}\!(\phi_0^{\rm cr}-\phi_0)}&{\phi_0^{\rm cr}-\phi_0\geq 1}\\
\end{array}} \right. \!\!\!.
\end{equation}
\noindent Note that the exponents for the MQT rates in
Eqs.~(\ref{eq:relax-rate}, \ref{eq:relax-rate2}) are not
proportional to the system's size or its total number of
particles. e.g. Ref.~\cite{ueda}. Indeed, while the effective
mass, e.g. Eq.~(\ref{eq:action-slow}), is proportional to $L$, the
height of the effective potential barrier scales as $L^{-1}$.

The value of parameter $\eta$ in Eqs.~(\ref{eq:relax-rate}, \ref{eq:relax-rate2}) for the experiments in
Ref.~\cite{Boshier} is $\sim 300$ and therefore fine tuning of
parameters is needed in order to obtain reasonable MQT rates. While this fact presents a formidable
difficulty in some MQT proposals, e.g. Ref.~\cite{ueda}, it seems that such fine tuning is relatively simple
for the present system. Indeed, the value of parameter $\phi_0$  is determined solely by the system's geometry
and by the frequency of rotation, but not by the density, and therefore can be controlled with desired precision.
Moreover, the value of $\eta$ can be reduced
by decreasing the cross-section of the ring ~\cite{comm6}, or by
increasing the scattering length, i.e., the value of $\lambda$,
via the Feshbach resonance.
\begin{figure}
\includegraphics[width=6.5cm]{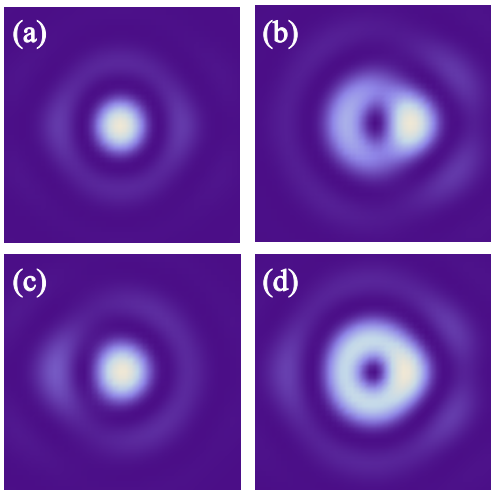}
\caption{(Color online) Calculated TOF measurement outcomes for: (i) Transitions
from the metastable current state $2\to 1$, Fig.~\ref{BEC.eps},
with $\Omega=0$. Observation of (a) corresponds to state ``1" and
(b)--- to state ``2". (ii) Transitions to the current-carrying
state in ($\Omega\neq 0$) case. Image (c) corresponds to state
``3" and (d) --- to state ``4".}\label{TOF.eps}
\end{figure}

Finally we briefly discuss the signatures of the current-carrying
metastable states in the Time-Of-Flight (TOF) measurements. As
well known, after the trapping potential is turned off, at
sufficiently long times of expansion the resulting BEC state
corresponds to the Fourier transform of the initial state in the
trap. In Fig. 2 we present numerical Fourier transforms of the
initial states that have same particle density $|\psi_0(r,
\theta)|^2$ shown in Fig. 1(a), but multiplied by different phase
factors $\exp{(i n\theta)}$, where phases $2\pi n$ correspond to
the values of $\phi$ in minima $1$ through $4$ of the effective
potentials in Fig. 1(b) and Fig. 1(c). That is, $n=0$ for the
state $1$, while $n\simeq 0.75$ for the state $2$, and so on. The
main distinction between Fourier images of states with $n\sim 0$
and with $n\sim 1$ is clear: While the former case corresponds to
the momentum distributions mainly concentrated at the origin, e.g.
Figs. 2(a) and 2(c), in the later case the distribution is {\it
ring-like}, e.g. Figs. 2(b) and 2(d) \cite{comm7}.

In summary, we have proposed a novel BEC based qubit-like device and
theoretically studied its properties. We have identified a set of
macroscopic metastable states that such
system is expected to exhibit under certain conditions and
evaluated transition rates between these states. We have discussed
signatures of these states in the standard TOF measurements and argue
that the effects considered in this paper can be readily observed
in contemporary cold atom systems.

{\acknowledgements We thank I. Martin, V. Privman and E.
Timmermans for valuable discussions and comments. The work is
supported by the US DOE. }


\end{document}